# The quenched $g_A$ puzzle in nuclei and nuclear matter and "pseudo-conformality" in QCD

Mannque Rho

*Institut de Physique Théorique, Université Paris-Saclay, CNRS,
CEA, Gif-sur-Yvette 91191, France
mannque.rho@ipht.fr*

The long-standing puzzle of the quenched $g_A$ in nuclei is shown to have an extremely simple resolution in a renormalization-group (RG) treatment of a hidden local symmetric (HLS) and scale-symmetric (HSS) chiral Lagrangian. It is shown that the Landau–Migdal fixed-point approximation in nuclear matter (or $V_{\rm lowk}$ in finite nuclei) in RG approach to strong correlations of fermionic hadrons on the Fermi surface *exactly* reproduces the superallowed Gamow–Teller transitions in the "Extreme Single-Particle (shell-)Model (ESPM)" in doubly-magic closed shell nuclei. One arrives at the quenching factor $q \approx 0.78$ giving the quenched $g_A^{\rm eff} \approx 1$. This resolution exposes scale-chiral symmetry, hidden in QCD in the vacuum, emerging in nuclear matter from low density to high compact-star density. It has important implications on "first principles" approaches to nuclear physics, such as the role of multi-body exchange currents in weak axial-current matrix elements in nuclei and in neutrinoless double $\beta$ decays for going Beyond the Standard Model. This resolution could put in serious doubt the most recent improved measurement of the superallowed Gamow–Teller transition in the doubly-magic closed shell nucleus $^{100}$Sn which if confirmed would require a "*fundamental quenching*" $q_{\rm ssb} \sim 1/2$.

## 1. Introduction

### 1.1. *The puzzle*

There is a long-standing "puzzle" lasting more than four decades as to why the Gamow–Teller transition in shell model in light nuclei with baryon number $A \lesssim 20$ requires what appears to be a "universal" quenching factor $q \sim 0.75$–$0.78$ multiplying the axial coupling constant $g_A$ measured in neutron decay. This quenching factor would make the effective axial-vector coupling constant $g_A^{\rm eff} \approx 1$ in nuclei.

M. Rho

What was striking then — and is more mysterious now — is that the resulting $g_A$ is surprisingly close to 1 in light as well as medium heavy nuclei as updated in the recent review.[1] This prompted Denys Wilkinson already from early 1970s[2] to inquire whether this is associated with something intrinsically tied to a basic property of the non-Abelian gauge theory QCD, perhaps not encoded in the matter-free vacuum but *specifically in the nuclear medium*. While the conserved vector current implies that the vector coupling constant $g_V = 1$ in and out of medium, the conserved axial current — inside the chiral limit, i.e. massless quark limit — does not imply that $g_A = 1$. This is now understood as that the axial symmetry is a "hidden" symmetry unlike the vector symmetry which is unhidden. This hidden symmetry nature has been an unproven "hunch" in nuclear community in 1970s that $g_A^{\rm eff}/g_A = 1 + (2 \pm 6)$ with 1 standing for the "full mixing" in shell model with the spread $2 \pm 6$ coming from higher order correlations in the wave function,[2] including possible contributions from $\Delta$ resonance components.

I should stress that there is, up to date, no clearly known and acceptable answer as to whether the apparent nuclear $g_A^{\rm eff} \approx 1$ is not just a coincidence in Nature. And if indeed it is not, what it means. In this paper, I attempt to provide a plausible explanation built on an old resolution of the problem I have been working on and off since the Les Houches Lectures.[2] But in arriving at a possible understanding, I encounter further "mysterious" issues arising, more or less totally ignored, not just in low-energy nuclear physics but also in the properties of hadronic matter under extreme conditions met in compact-star physics — and also in cosmology. The purpose of this note is to arrive as concisely as possible at what could be happening, which involves even the currently controversial issue of scale (conformal) symmetry in QCD with an IR fixed point for the number of flavors $N_f$ near 2 or 3 relevant to nuclear physics.

To see briefly what the problem is, consider the celebrated Adler–Weisberger sum rule that follows from the current algebras of chiral symmetry,[3,4]

$$g_A^2 = 1 + f_\pi^2 \frac{2}{\pi} \int_{m_N+m_\pi}^{\infty} \frac{WdW}{W^2 - m_N^2} [\sigma^{\pi^+ p}(W) - \sigma^{\pi^- p}(W)].$$

Applying this sum rule for the proton naively to a nucleus $A$ treated as an "elementary particle"[5] of the same flavor-global quantum numbers as the proton, one notes that $g_A \to 1$ if either $f_\pi \to 0$ or the integral over the difference of $\pi^\pm A$ scattering vanishes unless there are some corrections applicable in nuclear medium that go beyond the current algebra relations. There is nothing to suggest that the second term should vanish unless $f_\pi \to 0$. It is believed that the pion decay constant in medium will indeed go to zero — in the chiral limit — at some high density independently of how the second term behaves. In finite nuclei, however, in the vicinity of the equilibrium nuclear matter density $n_0 \simeq 0.16$ fm$^{-1}$, the in-medium $f_\pi^*$ could drop, at most $\sim 20\%$, from the free space value and there is no known reason why the integral is to vanish. Thus the puzzle of the effective $g_A^{\rm eff}$ tending to go near 1 in a wide range of nuclei.

### 1.2. *"First-principles" approach to nuclear physics*

This issue got highlighted recently in the nuclear physics community by a "work-of-the-art" approach for what is heralded as "first principles" computation of Gamow–Teller transitions in light and medium nuclei, in particular in $^{100}$Sn.[6] To correctly interpret the result of first-principles approaches in nuclear physics, one would have to clearly specify what one deals with in addressing the $g_A^{\rm eff}$ problem. From the point of QCD, admittedly the correct theory of the strong interactions, one first needs to define at what (energy) scale it is to be applicable as an effective theory (EFT). In nuclear physics, one is dealing with the "chiral scale" $\Lambda \sim 4\pi f_\pi \sim 1$ GeV, so the appropriate EFT has the degrees of freedom integrated out (modulo quantum anomalies) above the cut-off at $\Lambda$. For low-energy processes, the cutoff scale is effectively lowered to $\lesssim m_V$, the lowest vector meson mass $\sim$600–700 MeV. The currently favored — and highly successful — effective field theory is the standard S$\chi$EFT$_\pi$ of Weinberg anchored on soft-pion interactions involving the proton, neutron and the pion field $\pi$ as the relevant fields with the effective cutoff at $\Lambda^{\rm eff} \sim 200$–300 MeV. Weinberg called this approach "Folk Theorem" on EFT.[7] In the meson sector, the vector mesons are integrated out, so only the pion is taken into account together with the nucleons interacting in "soft-pion" kinematics.

Now the well-known fact in low-energy nuclear processes is that the nucleon with its mass $\sim$1 GeV has to be treated in a heavy-baryon formalism so that "soft-pion" kinematics figures in the chiral power counting. I will come back to the question as to what could be missing in this heavy-baryon formalism at high density.

The vector mesons are integrated out as mentioned. But what about scalar degrees of freedom? In phenomenological Lagrangian approaches to nuclear interactions, a scalar meson in the literature plays a crucial role for binding nuclear matter — as well as finite nuclei. The appropriate mass scale for such a role is $\sim$500–600 MeV. The scalar meson listed in the particle data booklet, $f_0(500)$, may be interpreted as a candidate for a pseudo-Nambu–Goldstone boson of spontaneously broken scale (or conformal) symmetry. The effect of the scalar meson with a mass greater than that of the pion could be taken into account in the higher order terms as for the vector meson in the systematic chiral expansion.[a]

The "work-of-the-art" computation[6] adopts this S$\chi$EFT$_\pi$ strategy. The "no-core shell model" technique exploited therein is claimed to capture "virtually exact" correlations — in the sense of Ref. 2 — in the nuclear wavefunctions in effective field theory (EFT) treatment of strong and weak interactions of the Standard Model. Calculations along similar lines in shell model in light nuclei have been around since $\sim$1980, but what distinguishes this work[6] from the previous works is the accuracy with which both high-order nuclear correlations and effective field theory treatment of nuclear force and many-body weak currents could *in principle* be put together in heavy nuclei. Furthermore what makes this work noticeable is that it is focused on the *superallowed* — $q/\omega$ (momentum/energy transfer) $\to 0$ — Gamow–Teller (GT) decay of the doubly magic $^{100}$Sn nucleus that exhibits the strongest Gamow–Teller strength so far measured in nuclei,[8] an ideal system for large-scale calculation that can take into account a large number of particle–hole correlations. It predicts in the sophisticated no-core shell-model technique combined with the nuclear force and weak currents including many-body currents treated in S$\chi$EFT$_\pi$ "the quenching" factor $q = 0.73$–$0.85$ for $^{100}$Sn, agreeing with $q_{B_{\rm GT,ESPM}} = 0.75(2)$ obtained in Ref. [8], thereby giving $g_A^{\rm eff} = 0.95$–$1.08$". This calculation is offered as a resolution of the long-standing puzzle.

---

[a]Note that even the pion can be integrated out leading to the pionless EFT at very low energy.

### 1.3. *The resolution?*

But there are serious caveats, I assert, to this resolution.

The first is that it ignores possible fundamental QCD sources for quenching that result from the cutoff scale picked for EFT. One prominent element is the trace anomaly associated with a dilaton — denoted as $\sigma$ — resulting from the spontaneous breaking of scale (or conformal) symmetry of QCD. This can lead to what I shall call *fundamental quenching* (*FQ*) of $g_A$ not taken into account in Ref. [6].

The second is the Brown–Rho scaling associated with the dilaton condensate $\langle \sigma \rangle$ in nuclear medium sliding in density, capturing the vacuum change by nuclear matter.[b] This effect is reflected in how the nuclear tensor forces — emerging from the integrated-out hidden local symmetric fields — control the Gamow–Teller matrix elements in the vacuum sliding in density. This is an important feature of the GT transition not accounted for in S$\chi$EFT$_\pi$.

The mechanism that is principally exploited in Ref. [6] is the effect of 2-body exchange current resulting from the heavy iso-vector mesons (e.g. $\rho$) integrated out. In what follows, such heavy mesons will naturally emerge from hidden local symmetry fields dual to QCD fields when the energy–momentum scale is extended higher. This will be described below.

In S$\chi$EFT$_\pi$, the many-body exchange current (mBC) effects are encoded in the higher derivative terms in the Weinberg chiral expansion and the way they enter in the weak axial current matrix elements is dictated by what was referred to as "chiral filtering mechanism"[9] which will be described below. Applied to the space component of the nucleon axial current relevant to the (superallowed) Gamow–Teller transition, the next-to-leading order (NLO) 2BC (and 3BC) terms, relative to the leading-order in the power counting for the single-particle term — which is super-allowed — come at N$^\kappa$LO with $\kappa \geq 2$ for small momentum transfer $q \lesssim m_\pi$. This means that the correction coming at high orders in $q \ll m_\pi$ — the heavy-meson exchange as well as the heavy-baryon formalism mentioned above — must be highly suppressed within the framework of the Folk Theorem. This in fact is in stark contrast to the axial-charge transition for which $\kappa = 0$ in all multi-body terms (including the possible corrections due to the treatment of baryons in heavy-baryon formalism). This follows from the soft-pion theorem associated with the "chiral filtering" with chiral symmetry described below.

---

[b]Let me mention here what has been wrongly understood in nuclear physics community. In the past, most of the heavy-ion community discussing dilepton production at high temperature totally mistakenly identified the dilaton condensate $\langle \sigma \rangle \sim f_\chi$ with the quark condensate $\langle \bar{q}q \rangle \sim f_\pi$. This relation may not be entirely incorrect at least numerically at low temperatures, but it is definitely wrong at temperatures near the chiral transition at high temperature. It is known that the dilaton decay constant behaves quite differently from the pion decay constant at increasing density, particularly near the chiral symmetry restoration density. This is because chiral symmetry and scale/conformal symmetry differ near the IR fixed point of scale symmetry in QCD for small number of flavors.

Now in Ref. 6, by playing with the "resolution scale" allowed in the effective theory, the quenching is "made" to shift from the leading single-particle GT operator to the 2-body exchange-current operator (2BC), and arrive at the quenching factor $q \approx 0.75$ leading to $g_A^{\text{eff}} \approx 1$ in the superallowed GT transition in $^{100}$Sn nucleus.

Such a manipulation is at odds with the Folk Theorem with the specified counting rule. Furthermore some 2BC terms — such as the "recoil terms"[10] belonging to the same power counting stemming from non-relativistic approximations — are missing. Hence the appropriate in-medium Ward identity cannot be satisfied. More importantly at N$^{>2}$LO, there are simply too many unknown parameters, e.g. $\gtrsim 11$,[11,12] in the Lagrangian to be fixed. Now if the incomplete sum of N$^2$LO terms are non-negligible, there is no justification to ignore higher-order 2BC as well as nBC terms with $n > 2$.

It should be mentioned that an effect of similar nature associated with the resolution scale was already encountered in 1970s. Suppose the cutoff scale is picked so that $\Delta$ resonance enters into the EFT. Then applying the connection — via PCAC — to pion–nuclear interactions, it has been shown[13] in Landau–Migdal Fermi-liquid formalism — that will be explained below — that the entire quenching effect can be moved from the particle-hole couplings to the $\Delta$-hole coupling $g'_{\Delta N}$ in the fixed-point 4-Fermi interaction parameters. The sum of the $\Delta$-hole bubble diagrams in the pion–nuclear optical potential inserted into the GT amplitudes has been found to lead to $q_{\text{quench}} = 1/(1 + \alpha)$ where $\alpha$ is related to the Ericson–Ericson–Lorenz–Lorentz (EELL) effect in $\pi$-nucleus scattering.[14] The result at nuclear matter density $n_0$ gives[13] $q_{\text{quench}} \approx 0.76$, essentially the same as in Ref. 6. Now the caveat here is that the Landau–Migdal parameter that figures here as in the pion–nuclear optical potential is that the universality $g'_0 \Delta N = g'_0 NN$ should be satisfied. Moving the quenching by fiat to the $\Delta$-hole as done in Ref. 13 was not justifiable within the framework of S$\chi$EFT$_\pi$. I should, however, mention that when the EELL effect, su̅itably combined with the Brown–Rho scaling, could lead to an extremely interesting justifiable possibility. It has not yet been worked out in a consistent way for the case of the quenched $g_A$. Such a possibility was in fact discussed in a different but related context by G. E. Brown in an unpublished paper quoted in Ref. 15. The crux of the matter there as stated by Gerry in referring to Landau–Migdal Fermi liquid theory is that "it is magical how EELL runs the show and makes, together with Brown–Rho scaling, all forces equal". In what follows, the approach G$n$EFT — to be defined below — is, I assert, to capture the essence of Gerry's idea, though phrased in a different language, and could lead to the resolution of the problem.

Finally, the most pertinent issue for the future via-à-vis with the quenched $g_A$ problem is the possible presence of a substantial "fundamental quenching $FQ$ factor" denoted $q_{\rm ssb} < 1$ revealed in the most recent improved RHIC measurement. This effect, as far as I am aware, has not been addressed in the literature, certainly not in Ref. 6.

## 2. Resolving the Quenched $g_A$ Puzzle

I first give a brief account of what's involved, followed by the principal results of this paper. Most of the arguments were given in sufficient detail in Refs. 16 and 17. I will offer very brief explanations of what is obtained, referring the details to the appropriate publications. What matters will be explained in as simple a way as possible.

### 2.1. *The model: GnEFT*

The process I will more or less entirely focus on is the superallowed GT transition in the doubly magic closed-shell nucleus $^{100}$Sn treated in Ref. 6. This process is found to present the condition where the arguments developed can be best met in the effective field theory termed as G$n$EFT. The model, G$n$EFT, goes beyond the standard chiral EFT, with the cutoff scale put at the vector meson mass scale $\gtrsim O(m_\omega)$, thus at higher scale than that of S$\chi$EFT$_\pi$. It consists of the usual pion field $\pi$ and the baryon field coupled to the hidden local symmetry (HLS) vector fields $\mathcal{V}_\mu = (\rho_\mu, \omega_\mu)$ together with the dilaton field $\sigma$ with an IR fixed point with the conformal compensator field for the dilaton $\sigma^{\rm c}$

$$\bar{\chi} = f_{\bar{\chi}} e^{\sigma/f_{\bar{\chi}}}, \tag{1}$$

where $f_{\bar{\chi}}$ encodes the dilaton condensate with hidden scale symmetry (HSS). The effective Lagrangian constructed with both HSS and HLS implemented to be applicable from normal nuclear matter $n_0$ to compact-star matter density $n_{\rm star}$ will be denoted as $\mathcal{L}_{\psi\sigma HLS}$. It generalizes S$\chi$EFT$_\pi$ as explained to a large range of densities up to the putative density at which the gravitational collapse is supposed to take place.

### 2.2. *The main results*

Let me start by giving the principal results and then explain them afterward.

Taking into account the quantum anomaly giving the quenching factor $FQ$ denoted $q_{\rm ssb}$, the leading scale-chiral-order nuclear axial current is given by

$$J^{\pm}_{\mu 5} = q_{\rm ssb} g_A \bar{\psi} \tau^{\pm} \gamma_\mu \gamma_5 \psi \tag{2}$$

---

$^{\rm c}$In what follows, unless confusing otherwise with chiral symmetry, I will use $\chi$ without bar to represent the linearly transforming (scale) conformal compensator field.

with

$$q_{\rm ssb} = c_A + (1-c_A)\Phi^{\beta'}, \qquad (3)$$

where $\beta'$ is the anomalous dimension of the energy–momentum tensor $\text{Tr}(G_{\mu\nu})^2$, $\Phi$ is the Brown–Rho scaling $\Phi = f_\chi^*/f_\chi$ and $0 \leq c_A \leq 1$ is an arbitrary constant. The $q_{\rm ssb}$ is the fundamental quenching ($FQ$) inherited from the scale/conformal anomaly of QCD proper. How this comes about will be explained later. The possible contribution $q_{\rm ssb} < 1$ in the current is the new development, absent up to date in all publications in the field of nuclear/astrophysics, including Ref. 6.

The first of the two principal results of this paper is that the most recent experimental indication in superallowed Gamow–Teller transitions in medium heavy nuclei indicates that $q_{\rm ssb}$ could be considerably smaller than 1 in nuclear matter

$$q_{\rm ssb}(n \approx n_0) \approx 0.5. \qquad (4)$$

Such a big quenching factor, if confirmed, could impact extremely importantly, among others, the search for going beyond the Standard Model in $0\nu\beta\beta$ decays in nuclei. It would also bring paradigm change in nuclear theory.[d]

The second of the issues is that nuclear correlations are governed by an emerging scale symmetry in nuclear medium by what is referred to as "pseudo-conformal symmetry" which accounts for $g_A q_{\rm snc} \approx 1$ ($q_{\rm snc}$ is the "strong nuclear correlation" quenching in finite nuclei) in nuclear matter, dense compact star matter, etc. up to the dilaton fixed-point density $q_{DFP} \gtrsim 20n_0$ where $\langle\chi\rangle \to 0$. This also could bring a drastic change in nuclear theory.

In what follows how these quantities are obtained will be explained as concisely as possible.

### 2.3. $S\chi EFT_\pi$

Let me start by what is in the standard nuclear EFT, $S\chi\text{EFT}_\pi$ and what could be missing in it — apart from the $FQ$ — for the quenched $g_A$. Treating nucleons in terms of heavy-baryon formalism as appropriate for soft interaction for $S\chi\text{EFT}$,[e] to account for the degrees of freedom integrated out at $\Lambda_{\rm eff} \gtrsim 200$–$300$ MeV, say, HLS and HSS, one makes a power expansion in $O(p) \sim O(\partial) \sim O(m_\pi)$ as counter terms in the effective Lagrangian. This necessitates for nuclear processes 2- and 3-body interactions between nucleons and 2- and 3-body exchange currents for nuclear EW responses.

---

[d] As mentioned at repetition, the coupling $g_A$ figures in nuclear interactions in the pion–nuclear coupling via the Goldberger–Treiman relation, hence in $S\chi\text{EFT}_\pi$.
[e] This procedure accounts for possible defects of $S\chi\text{EFT}_\pi$ at high density. If one integrates out the nucleons also since $m_N \gg \Lambda_{\rm eff}$, they should reappear as skyrmions so as to capture high density physics. This brings out an interesting aspect in dense matter physics in terms of 1/2-skyrmions as is discussed elsewhere.[16]

*M. Rho*

### 2.3.1. *Many-body currents*

In applying S$\chi$EFT$_\pi$ to nuclear processes, the nuclear forces $V_{\mathbf{nucl}}$ and the many-body currents (mBC) are considered, respectively, up to N$^4$LO and N$^3$LO in the given chiral power counting.[f] Let us assume that with a highly reliable $V_{\text{nucl}}$ the wavefunctions could be *accurately calculated* with the given powerful quantum many-body techniques. This is considered as the fist step to what might be called a "first-principles" calculation. There can of course be an objection here given that the shell model is not derived in S$\chi$EFT$_\pi$.[g]

Now the issue of the many-body currents (mBC) needs to be addressed.

It turns out that unlike the vector currents that are quite straightforward the nuclear axial-vector currents turn out to be extremely subtle. This has to do with that the axial symmetry is "hidden".[h]

In fact it has been known since late 1970s that the space and time components of nuclear axial currents behave quite differently in nuclei and dense baryonic matter. This was evidenced in the current algebras before the advent of QCD in the way "soft pions" come into two-body exchange currents.[9] In terms of the modern $\chi$EFT parlance, this is almost trivial. However the soft-pion theorems, just as all other soft theorems, be that photon or graviton, have a deep physical implication, ubiquitous in all areas of physics.[18]

### 2.3.2. *Soft theorems and the chiral filter: Story of two sides of a coin*

A simple heuristic picture of what is going on is as follows.

Consider the two-body axial currents with one-pion exchange which in the S$\chi$EFT$_\pi$ are dominant in the chiral power counting as we will see precisely below. Since the axial field $\mathcal{A}_\mu$ does not couple to the pion exchanged between nucleons, what is involved is the vertex $\mathcal{A}_\mu + N \to \pi + N$. We can take the axial field $\mathcal{A}_\mu$ as a pion. Thus we are dealing with the process $\pi_{\text{in}} + N \to \pi_{\text{out}} + N$ where $\pi_{\text{in}}$ stands for the incoming axial field and $\pi_{\text{out}}$ is the pion exchanged between two nucleons.

First consider the case where $\pi_{\text{in}}$ is "hard" and $\pi_{\text{out}}$ is "soft". Then according to the double soft theorems,[18] the amplitude should be highly *suppressed* by Adler's theorem. On the other hand, if both $\pi_{\text{in}}$ and $\pi_{\text{out}}$ are soft, then the double-soft limit gives an unsuppressed $\sim O(1)$ amplitude. This is very well known from the old soft-pion theorems, but nowadays this old stuff has become highlighted because of its fundamental nature in physics.

---

[f]There may be some differences in power counting from other authors. In this paper, I use the power counting *relative* to the leading order (LO) in the chiral expansion parameter $Q \sim p \sim \partial \sim m_\pi$ as follows: If the leading order (LO) term is of $O(Q^k)$, the subleading term of $O(Q^{k+m})$ for $m > 0$ is denoted as N$^m$LO, etc.

[g]A strict consistency would require that given the EFT Lagrangian of QCD, the shell model be "derived" with the nuclear force given by the Lagrangian, say, as a sort of non-topological chiral soliton or something similar. There have been some attempts to do this but there has been little progress in that endeavor. For instance, getting Fermi surface in a system of interacting fermions is a quantum critical phenomenon and using the EFT Lagrangian for fluctuations on top of given Fermi surface is already a hybrid approach with inevitable disregard of strict consistency. The same goes with the shell model.

[h]The statement that it is "spontaneously broken" is a misnomer.

Kinematics of the virtual pion in nuclei is not sharply given, so our argument is at best approximate. But with the axial current identified with a pion, this soft-theorem can be applied to the problem. The pion exchanged between two nucleons favors the process when it is soft, with harder pions suffering from kinematic suppression due to the derivative coupling. Now taking the axial charge operator $\mathcal{A}_0$ as a soft pion, this then predicts an $O(1)$ one-pion exchange contribution whereas the Gamow–Teller operator $\mathcal{A}_\pm$, being "hard", leads to a suppressed two-body operator. This is essentially the "chiral filter" argument of Ref. 9.

### 2.3.3. *Chiral filtering in $S\chi EFT_\pi$*

The above chiral filter argument can be given a more rigorous support by using the systematic power counting in $S\chi EFT_\pi$ to which we turn.

The power counting for the responses to the electro-weak current, first worked out in early 1990s and listed completely in a publication in 2003,[11] has been extensively refined and extended since then as aptly summarized — with relevant references — in Ref. 12. What we need for our arguments is essentially all contained in Ref. 11 that we will follow. We will, however, exploit some of the critical analyses made in Ref. 12 on consistency in the regularization schemes in going to N$^4$LO.

What follows is a story of two sides of the same coin.

We first look at the time component of the axial current. Here soft pions predominantly enter in the two-body current. The ratio of the two-body soft-pion exchange operator over the one-body operator — which is $O(Q)$ in the $Q$-power counting — is $R = 2B/1B = O(Q^0)$. Thus the leading "correction" is of the same magnitude as the LO one-body term. The next correction is suppressed by two chiral orders, $O(Q^2)$. At this order there are relativistic and other small corrections to the single-particle operator as well as two-body terms involving $2\pi$ exchange, etc. They are expected to be ignorable. Thus the leading two-body term is protected by the "chiral filter", hence robust.

This prediction has been neatly confirmed in the first-forbidden $A$-to-$B$ nuclear $\beta$ decay $A(0^\pm) \to B(0^\mp) + e + \nu, \Delta T = 1$ where the superscripts are the parities. Expressed as $\epsilon = g_{A\,t}^{\text{eff}}/g_A$ in terms of the effective axial coupling constant for the time component to represent the ratio of the total matrix element over the single-particle matrix element, the prediction for nuclear matter[19] $\epsilon_{\text{theory}} = 2.0 \pm 0.2$ and the experimental measurements made for the transitions in Pb region $A = 205$–$212$[20] $\epsilon_{\text{exp}} = 2.01 \pm 0.05$ agree stunningly well. The theoretical value is estimated at nuclear matter density, but the result is extremely insensitive to density, so Pb can be compared with nuclear matter: The 10% error bar assigned to the theory corresponds to the range of density involved from light to heavy nuclei to nuclear matter. This result is well supported in other processes involving lighter nuclei. This perhaps

*M. Rho*

is the most convincing — and clear-cut — evidence for the role of soft pions — via exchange currents — in nuclear physics.

This is the story of one side of the coin.

Now look at the Gamow–Teller operator, the space component of the axial current. The situation here is drastically different. This is because soft pions play practically no role here. While the one-body Gamow–Teller operator is $O(Q^0)$, super-allowed barring accidental suppression, the leading two-body correction with one-pion exchange comes (à la soft theorems) strongly suppressed by two chiral orders, $O(Q^2)$. The ratio is $R = 2B/1B = O(Q^2)$. This is because the pion entering in the two-body term is "hard", with its coupling with nucleons requiring, among others, relativistic corrections. At this order, three-body operators must also enter. Furthermore since the nucleons are inevitably non-relativistic, there can be a plethora of other corrections, notably the "recoil corrections"[10] entering at the same order. Possible corrections to the heavy-baryon formalism employed in S$\chi$EFT$_\pi$ must enter at this chiral order. It does not appear from what is discussed in Ref. [6] that all these corrections are fully — and consistently — taken into account. There is no justification to take some terms but ignore others as there can be significant cancellations among them. There should also be requirements for axial Ward identities. To make it even worse, there are also ambiguities in doing regularizations in both $V_{\rm nucl}$ and mBO,[12] relevant to the validity in correlating the presence of 2BC with the regularization ("resolution scale", etc.) that figures importantly. This large number (>11) of higher-order terms that cannot be controlled at N$^k$LO for $k \geq 3$ is what is meant by "chiral-filter unprotected" terms including in particular those effects coming from what is absent in the baryon structure captured in the heavy-baryon formalism.

This is the story of the other side of the same coin.

In sum, we assert that the conclusion[6] — that the $g_A$ problem is resolved by the 2BC combined with a sophisticated no-core shell model — is highly questionable. There is no justification to stop at N$^2$LO unless N$^3$LO can be shown to be ignorable, a task which is at present far from feasible. There can very well be cancelations between hidden different orders as in the case of the Monte Carlo calculations in light nuclei.[21]

### 2.4. *Hidden scale-chiral HLS symmetric EFT: GnEFT*

Let me now present the approach that provides a strong support to the conclusion given above. This calculation relies on the Lagrangian $\mathcal{L}_{\psi\sigma\rm HLS}$ mentioned above. As reviewed in Ref. [16], it was formulated to properly account for the properties of baryonic matter from near the nuclear matter density $n_0$ to high densities relevant to massive compact-star matter $\gtrsim 3n_0$. This means starting from the Lagrangian $\mathcal{L}_{\psi\sigma\rm HLS}$, makes it work from $n_0$, going beyond the putative hadron-quark cross-over density $n_{\rm HQ} \gtrsim 3n_0$ and confronting compact-star densities $\sim$5–7$n_0$. The strategy is built on Landau–Migdal Fermi-liquid theory treating, in renormalization-group technique, strongly-correlated Fermions on the Fermi sea as in condensed matter physics but incorporating the hidden symmetries involving the pions, the vector mesons and the dilaton $\sigma$.[i] This means that the starting Lagrangian is $\mathcal{L}_{\psi\sigma\text{HLS}}$. Though it is rather involved including a possible hadron-quark continuity at high density, it drastically simplifies for densities $\leq n_0$, so it is neatly applicable to the problem concerned. It involves essentially only one parameter, namely, the pion decay constant in medium $f_\pi^*$ which can be extracted from experiments.

### 2.4.1. *The Lagrangian $\mathcal{L}_{\psi\sigma\text{HLS}}$ with an IR fixed point*

The Lagrangian $\mathcal{L}_{\psi\sigma\text{HLS}}$ is constructed with a cutoff put above the vector-meson mass ∼700 MeV, with the vector mesons $\rho$ and $\omega$ brought in as hidden gauge fields and a scalar dilaton field $\sigma$ which although not proven yet, could be identified with $f_0(500)$. The vector mesons are treated as the hidden local symmetric (HLS) bosons. It is more or less standard by now in nuclear physics with even stronger support in recent developments.[22,j] How to implement scale/conformal symmetry, however, is a highly subtle and controversial matter in QCD — and other areas of physics — since 1970s and still is. I will adopt the recent development which indicates that there can be an IR fixed point which can accommodate a dilaton for QCD with flavor number ≤3 with a free-space mass comparable to that of the kaon mass. It has worked fairly well so far without any fatal inconsistency with the presently known properties of nuclear matter as well as of compact-star matter.[16,17]

Written schematically, the Lagrangian is of the form

$$\mathcal{L}_{\psi\sigma\text{HLS}} = \mathcal{L}_{\text{inv}}(\psi, U, \chi, V_\mu) + \mathcal{V}(U, \chi, \mathcal{M}), \quad (5)$$

where the first term is scale-invariant and the second is the dilaton potential that encodes scale-chiral symmetry breaking. This results from the leading-order scale symmetry approximation found to be appropriate for nuclear dynamics.[16] In particular, it encodes soft theorems. Here $\psi$ is the nucleon field, $U = e^{2i\pi/f_\pi}$ is the chiral field, $\chi = f_\chi e^{\sigma/f_\chi}$ is the "conformal compensator field" for the dilaton $\sigma$, $V_\mu$ is the hidden gauge field and $\mathcal{M}$ is the quark mass term. The dilaton potential encodes the vacuum structure putting the system in the Nambu–Goldstone mode of scale-chiral symmetry. The HLS is assured with hidden gauge covariance put in the Maurer–Cartan 1-forms and can be written down to any power orders. The key assumption made for the dilaton $\sigma$ is that there is an infrared (IR) fixed point for $N_f \leq 3$ relevant for nuclear processes. In the chiral limit, the $\beta$ function vanishes with both chiral symmetry and scale symmetry realized in Nambu–Goldstone mode with both $f_\pi$ and $f_\chi$ non-zero. At the IR fixed-point, they accommodate massive hadrons, e.g. baryons, vector mesons, etc.

---

[i] "Migdal" is appended to the Landau Fermi liquid theory given the additional strong-interaction degrees of freedom involved.

[j] It has been suggested that these HLS fields are in fact emergent dynamically generated boson fields with the properties assumed thus far, such as VD, KFSR with $a = 2$, etc.[23]

There are two versions for the dilaton, one called "genuine dilaton (GD)"[24] and the other "QCD-conformal dilaton (QCD-CD)".[25] Although the two versions appear to have some differences in detail, they are related, modulo parameters that figure in the application made in Ref. 16. The GD scheme was exploited in the early development[26] and will be followed in this discussion.[k]

### 2.4.2. *Quenching of $g_A$ in Landau–Migdal Fermi-liquid fixed-point approximation*

Given the Lagrangian $\mathcal{L}_{\psi\sigma\text{HLS}}$, the question to address is how to accurately calculate the superallowed Gamow–Teller matrix element in nuclear matter.

Consider specifically the GT transition in a medium heavy nucleus, say, $^{100}$Sn with the axial current (2). Let us for the moment ignore the $FQ$ and take $q_{\text{ssb}} = 1$. Given that it is a multiplicative constant, it can be readily reinstated later. This problem is best approached — for the case involved — by the Wilsonian renormalization-group approach to strong correlations of fermions on the Fermi surface developed in condensed matter physics.[27,28] The idea is to start with an effective Lagrangian modeling QCD at low energy and formulate it into a Landau's Fermi-liquid theory.[29] This was already done in 1996 following the strategy of going from a chiral Lagrangian to Landau parameters.[30] One can do this in two equivalent ways invoking "double decimations".[31] One can work with the HLS–HSS-implemented Lagrangian or in terms of four-Fermi marginal interactions. In the mean-field, one arrives at the Fermi-liquid fixed-point (FLFP) approximation defined as the limit $\bar{N} = k_F/(\Lambda_F - k_F) \to \infty$ where $\Lambda_F$ is the cutoff on top of the Fermi surface measured from the origin of the sphere.[27] To nuclear theorists, this procedure is familiar with the Walecka's relativistic mean-field approach[32] which leads to none other than the Fermi-liquid theory in nuclei.[33] Note also that this approach is consistent with Hohenberg–Kohn's density functional theory. The major difference of this approach anchored on $\mathcal{L}_{\psi\sigma\text{HLS}}$ from Walecka's mean-field approach is the important sliding vacuum structure due to the dilaton condensate $\langle\chi\rangle \propto \langle\bar{q}q\rangle^*$.

Before going into the principal problem of this paper, let me recapitulate how the approach works in nuclear EM response functions.[30] Here the situation is straightforward. An illustrative case is the EM orbital current in the mean-field treatment which reproduces precisely Migdal's finite Fermi-liquid formula[34]

$$\vec{J} = \frac{\vec{k}}{m_N}\left(\frac{1+\tau_3}{2} + \delta g_l\right),$$
$$\delta g_l = \frac{1}{6}(\tilde{F}'_1 - \tilde{F}_1)\tau_3,$$

where $\tilde{F}_1$ and $\tilde{F}'_1$ are Landau–Migdal interaction parameters expressed in terms of the parameters of the Lagrangian $\mathcal{L}_{\psi\sigma\text{HLS}}$.

---

[k]What is important to point out in advance is that a crucial issue of the $FQ$ arises from the specific prediction of the anomalous dimension $\beta'_{\text{IR}} = 0$ coming from Ref. 25.

There are two remarkable results in this formula. First the orbital current is given in terms of the vacuum nucleon mass — instead of the Landau mass ($m_L$) or the BR-scaled mass $m_N^* = \Phi m_N$ — satisfying the Kohn theorem[35] due to the back-flow diagrams required by the Ward identity and the other is that the prediction for the nuclear anomalous gyromagnetic ratio[30] — with the soft-pion theorems playing the crucial role — $\delta g_l^p(n_0) \simeq 0.21$ agrees with what is measured in the Pb region, $\delta g_l^{\text{proton}} = 0.23 \pm 0.03$.[36] As far as I know, this quantity has not been, up to date, explained by S$\chi$EFT$_\pi$.

Let me now apply the same Fermi-liquid fixed-point approximation to the superallowed Gamow–Teller transition of the quasiparticle sitting on top of the Fermi surface. This calculation was already done in Ref. 30. Expressed in terms of the LFL fixed-point quantities, the result was found (for $q_{\text{ssb}} = 1$)

$$q_{\text{snc}}^L \equiv [g_A^L/g_A] \approx \left(1 - \frac{1}{3}\Phi \tilde{F}_1^\pi\right)^{-2}, \qquad (6)$$

where $g_A^L$ is the Landau fixed-point constant valid for $\bar{N} \to \infty$ represented by the superscript $L$, $\tilde{F}_1^\pi$ is the pion Fock term contribution to the Landau parameter $\tilde{F}_1$ that enters into $\delta g_l$ and $q_{\text{snc}}^L$ is the complete nuclear correlation effect multiplying the single quasiparticle matrix element defined precisely below. The Fock term is a loop contribution, so naively $O(1/\bar{N})$. But the pion being "soft", it plays an indispensable role as it does for the anomalous orbital gyromagnetic ratio $\delta g_l^p$. Since pionic properties are given by chiral dynamics, the pion contribution $\tilde{F}_1$ can be calculated almost exactly. Thus once $\Phi$ — the only parameter in the theory — is given, then $g_A^L$ is accurately calculable. How the pion decay constant behaves in nuclear medium is experimentally measured,[37] so $\Phi$ is known in the vicinity of nuclear matter density. Quite surprisingly while $\Phi$ decreases as density increases, the pionic term $\tilde{F}_1^\pi$ increases with the product $\Phi \tilde{F}_1^\pi$ staying nearly constant as density changes, say, between $\sim \frac{1}{2} n_0$ and $\gtrsim n_0$. Therefore, $g_A^L$ is nearly density-independent, which predicts that the purely nuclear correlation quenching factor $q_{\text{snc}}^L$ must be more or less the same from light nuclei to heavy nuclei (and dense matter $n \gtrsim n_0$). The result at $n_0$ is

$$q_{\text{snc}}^L \simeq 0.78. \qquad (7)$$

This represents the single quasiparticle property that encodes the total nuclear correlation effects for the superallowed $q/\omega \to 0$ Gamow–Teller transition. This means that Eq. (7) is to capture the total nuclear quenching factor for the given transition in the effective field theory defined with the cutoff $\Lambda_{\text{chiral}}$ with the nucleons (p,n), HLS — $\rho$ and $\omega$ — and HSS dilaton — $\sigma$ — as the relevant degrees of freedom. What it captures in arriving at (6) is difficult to fully enumerate but in arriving at the simple formula, apart from the Brown–Rho (BR) scaling, certain symmetry properties of the constituent quark model valid at the large $N_c$ limit, and the Landau fixed-point approximation $1/\bar{N} \propto \Delta\epsilon/k_F \to 0$ (where $\Delta\epsilon$ is the single quasi-particle-quasi-hole excitation energy near the Fermi surface) are taken. Since in the large $N_c$ limit, $g_A$ goes $\sim O(N_c)$, one can limit to $O(N_c)$ in computing $g_A^{\text{eff}} = q_{\text{snc}}^L g_A$ (ignoring the $FQ$, i.e. $q_{\text{ssb}} = 1$).

$$g_A^{\text{eff}} = q_{\text{snc}}^L g_A \simeq 0.78 \times 1.276 \simeq 1.0. \tag{8}$$

Given that (6) is insensitive to matter density for nuclear matter, one can take this applicable to the light nuclei listed in Ref. 1 as well as a prediction for heavy nuclei with $q_{\text{ssb}} = 1$.

As remarked in Ref. 30, Eq. (6) is a Goldberger–Treiman relation relating $g_A$ to $g_{\pi NN}$ for a quasiparticle on Fermi surface. As for free nucleon, it encodes the large-$N_c$ limit known to be a good approximation for the Goldberger–Treiman relation. That it holds for the axial coupling constant for the Gamow–Teller transition in nuclei as a Landau fixed-point quantity was an assumption made in Ref. 30. It followed from the skyrmion description for the nucleon[38] that $g_A$ is directly related to the Skyrme quartic term — which is $O(N_c)$ — that stabilizes the skyrmion. As far as I am aware, this relation has not been shown to hold well in the standard nuclear chiral perturbation theory S$\chi$EFT$_\pi$.

Let me just remark here that in a recent development on HLS as an *emergent* symmetry by Yamawaki,[23] it has been shown that HLS follows in non-perturbative dynamics of the large (Grassmannian) N limit of the nonlinear sigma model based on the Grassmannian manifold. Remarkably, *all the parameters* of HLS, such as the vector dominance (VD), KSRF relations and $\rho$ meson universality get fixed by loop expansion and agree with Nature.[^1] What appears also remarkable is that the dynamically generated kinetic energy term of the HLS Lagrangian so derived gives rise to the Skyrme quartic term and hence makes the skyrmion stable, identifiable with the nucleon. It suggests that one could arrive at a formula of Eq. (6)-type in a systematic higher-order expansion in the Grassmannian model.

Up to date, there have been no systematic power-expansion corrections of the type developed in S$\chi$EFT$_\pi$ to the Fermi-liquid fixed-point approximation. However there is a recent development in condensed matter physics to access $1/\bar{N}$ corrections by nonlinear bosonization of Fermi surfaces with the method of coadjoint orbits.[39] In the $g_A$ problem, the hidden-symmetry degrees of freedom bring in subtle interplays with the Fermi surface modes that need to be worked out. A preliminary simplified calculation seems to indicate that the $1/\bar{N}$ corrections could be negligible over the range of densities involved.[40]

### 2.4.3. *Dilaton limit fixed point*

The mean-field approximation result in G$n$EFT, Eq. (6), arrived at the LFL fixed-point with the Lagrangian $\mathcal{L}_{\psi\sigma\text{HLS}}$, insensitive to density near $n_0$, could account for the $g_A^{\text{eff}} \approx 1$ seen in light and medium–heavy nuclei with small corrections $1/\bar{N}$ corrections, but interestingly there is an indication that $g_A \approx 1$ continues, unchanged to very high density. To see this, make in $\mathcal{L}_{\psi\sigma HLS}$ the field redefinition $\mathcal{Z} = U\chi f_\pi/f_\chi$

---

[^1]: For instance, the parameter *a*, an arbitrary constant, had to be fixed to 2 in Ref. 22.

and let $\text{Tr}(\mathcal{Z}\mathcal{Z}^\dagger) \to 0$ in the LFL fixed-point approximation. This is what is referred to in the literature as the dilaton-limit fixed point[16,41] that may be arrived at some high density $\gg n_0$ but not necessarily at the restoration of scale symmetry. That there be no singularity in this limit sets the constraints[m]

$$g_v \to g_A \to 1 \quad \text{and} \quad f_\pi \to f_\chi. \tag{9}$$

It also constrains that there be parity doubling in the nucleon spectrum and that the $\rho$ decouple from the nucleons in medium before the onset of possible vector manifestation.[22,n]

One arrives at a surprising conclusion that the effective $g_A^{\text{eff}}$ continues to be $\approx 1$ beyond the nuclear matter density and possibly beyond massive compact-star density conjectured in Ref. [17].

### 2.4.4. *Bridging FLF to ESPM*

What remains to do is to address how to exploit the result (7) or (8) for what is taking place in superallowed Gamow–Teller transitions in heavy nuclei. What is needed is the nuclear system in shell model that makes the transition *corresponding as closely as possible to* the LFL fixed-point transition with $1/\bar{N} \to 0$. This would involve the suppression of single quasi-particle-quasi-hole bubble diagrams in the standard many-body approaches. At present, there are no reliable full *ab initio* "first-principles" numerical treatments of complex nuclei. The superallowed GT transition in heavy nuclei in the "Extreme Single Particle shell Model" (ESPM)[43] comes closest to the conditions required for making the bridge from the shell model to the LPL fixed-point treatment.

The quenching factor in the Fermi-liquid fixed point theory — ignoring the *FQ* that will be restored later — $q_{\text{scn}}^L$ is what corresponds to the Fermi-liquid fixed point constant that multiples the zero-momentum-transfer matrix element $\mathcal{M} = (\sum_i \tau_i \sigma_i)_{QP}$ for the quasi-particle on top of the Fermi surface making the GT transition, $M_{GT} = q_{\text{ssb}} q_{\text{snc}}^L g_A \mathcal{M}$. This should be compared with the experimental value of the quenching factor $q$ in $^{100}$Sn.

The relevant part of the Lagrangian (5) in medium is

$$\delta \mathcal{L}^* = q_{\text{ssb}} g_A \bar{\psi} \gamma^\mu \gamma_5 \tau_a \psi \mathcal{A}_\mu^a + \cdots \tag{10}$$

with $q_{\text{ssb}}$ given by (13) where $\mathcal{A}_\mu^a$ is the external axial field. As mentioned, the *FQ* ($q_{\text{ssb}}$) appears due to the in-medium Ward identity.

At the Fermi-liquid fixed point, the relevant quantities involved are the Landau mass $m_L$, the Landau interaction parameters $\tilde{F}_1$ and $\tilde{F}_1'$ and $\Phi = f_\chi^*/f_\chi$. Thus $q_{\text{snc}}^L$ must involve only these quantities. The calculation for $q_{\rm snc}^L$ was first done a long time ago,[30] which to our surprise is exactly reproduced — for $q_{\rm ssb} = 1$ — by the considerably more improved argument.[o]

---

[m] It has been pointed out before but is worth noting that $f_\chi \to f_\pi$ differentiates the IF fixed structure of the GD (and QCD-CD) from what is involved in dilaton EFT for $N_f \gtrsim 8$ discussed for conformal Higgs where $f_\pi \ll f_\chi$.[42] This feature could underlie the scale symmetry in the precocious pseudo-conformal sound speed of compact stars predicted by G$n$EFT.[17]

[n] This is another indication that the search for dropping mass of the $\rho$ meson as the signal for Brown–Rho scaling in dilepton experiments was totally misguided.

Recalling that the LFL fixed-point limit in the Fermi-liquid system obtained above, Eq. (7), corresponds to a Gamow–Teller transition undergoing on the Fermi surface with the kinematics $\omega \to 0$, $q/\omega \to 0$, we require that the quasi-particle–quasi-hole bubble contributions entering at higher orders in $1/\bar{N}$ in G$n$EFT be suppressed. I propose that this allows the LFL fixed-point limit in the Fermi-liquid system to be mapped *exactly* to the EPSM in doubly-magic-shell nuclei. The best case is offered by the GT transition in the $^{100}$Sn nucleus which has the proton and neutron shells completely filled at 50/50. An advantage of this nucleus is that it is the heaviest nucleus with the equal magic shells filled close in density to nuclear matter that has also been studied extensively both experimentally and theoretically.[8,44,45] Furthermore it is this transition where a substantial fundamental quenching seems to be seen in experiment.[45] The process involved is a pure superallowed GT transition of a proton ($\pi$) $\pi 0g_{9/2}$ in the completely filled orbital into a neutron ($\nu$) in the empty spin–orbit partner, the $\nu 0g_{7/2}$ orbital of $^{100}$In. This offers the structure of the daughter state that is of the pure $(\nu g_{7/2})$particle–$(\pi g_{9/2})$hole state to which the ESPM could be applied. The amplitude obtained by the square of the single-particle element, $(\langle(\nu)g_{7/2}|\tau^+\sigma|(\pi)g_{9/2}\rangle)^2$, is denoted by $\mathcal{B}_{\rm GT}^{\rm EPSM}$. For the EPSM (assumed to be applicable to the nucleus), the N$^{\geq 2}$LO multi-body terms[11] can be dropped by the chiral-filter reasoning.

We can now proceed to do the mapping for the $^{100}$Sn GT decay. Phrased in terms of $\mathcal{B}_{\rm GT}^{\rm EPSM}$, the Landau Fermi-liquid fixed point GT strength can be written in terms of the ESPM quantities and leads to the theoretical prediction for the $^{100}$Sn GT decay[46]

$$\mathcal{B}_{\rm GT}^{\rm FLtheory} = \mathcal{B}_{\rm GT}^{\rm EPSM}(q_{\rm ssb} q_{\rm snc}^L)^2 \approx 10.8 q_{\rm ssb}^2, \qquad (11)$$

where $q_{\rm snc}^L$ is identified as the full nuclear correlation effect given in the FLP theory (7). Here $\mathcal{B}_{\rm GT}^{\rm EPSM} = 16/9$ is the EPSM value for the superallowed matrix element squared.[43] Equation (11) is the prediction I am making that follows from the matching of the Fermi liquid to the doubly-magic shell structure. Given a well-measured experimental value $\mathcal{B}_{\rm GT}^{\rm exp}$, one could then extract the fundamental quenching factor $q_{\rm ssb}$ by equating (11) to $\mathcal{B}_{\rm GT}^{\rm exp}$.

### 2.4.5. *Experimental indication for FQ*

There are several recent experiments that could be analyzed for the possible role of *FQ*, neatly disentangled from complicated nuclear correlation effects. There have been other measurements in the literature on the same transition but let me focus on two most recent ones that best illustrate the points. It turns out that they give definitely conflicting information: One from GSI[8] which shows no signal for *FQ* and the other from RIKEN[45] which definitely shows a strong *FQ*.

---

[o] In this reference, an argument was made using the Skyrme model with the scale invariant quartic term with the coefficient $1/e^2 \sim O(N_c)$ which leads to $g_A \sim O(N_c)$. The result is the same as is obtained now with (5) in the mean field.

• **GSI:** The extracted strength of the $^{100}$Sn decay to the $1^+$ daughter state in $^{100}$In comes out to be $\mathcal{B}_{\rm GT}^{\rm exp} = 9.9^{+2.6}_{-2.5}$. Given the large error bars involved, let us take it to be $\mathcal{B}_{\rm GT}^{\rm GSI} \approx 10$. Then from (11), one can conclude that there is no sign for *FQ*, $q_{\rm ssb}^{\rm GSI} \approx 1$. This then gives the observed $g_A$, $g_A^{\rm eff} \approx 1$. A more detailed account of possible corrections to the pure single particle–hole dominance daughter state does not affect the conclusion. Let me mention that this is the result arrived at in Ref. 6, not the "improved" RIKEN result quoted below.

It seems fair to state that *this GSI result does offer a support for an exact correspondence between the LFL fixed-point result* (6) *and the ESPM result for the pure superallowed GT transition of a proton($\pi$) $\pi 0g_{9/2}$ in the completely filled orbital in $^{100}$Sn into a neutron($\nu$) in the empty spin–orbit partner, the $\nu 0g_{7/2}$ orbital of $^{100}$In with $q = q_{\rm snc}^{\rm ESPM} = 0.78$.*

• **RIKEN:** The RIKEN result, claimed to be an "improvement" over the GSI one with much smaller error bars, comes out to give an entirely different picture to the problem. The measured strength $\mathcal{B}_{\rm GT}^{\rm RIKEN} = 4.4^{+0.9}_{-0.7}$ leads to the *FQ* with the quenching factor in the range

$$q_{\rm ssb}^{\rm RIKEN} = 0.46 - 0.55 \tag{12}$$

giving $g_A^{\rm eff} \approx 0.6$–0.7.

I should stress that this drastic quenching, if proven to be confirmed by further scrutiny, would apply not just to superallowed GT transitions but also to *all* weak transitions involving nucleons in nuclear medium. It would clearly bring a serious problem in nuclear physics given that $g_A^{\rm eff}$ should figure in *all nuclear processes* involving pions via the Goldberger–Treiman relation.

### 2.4.6. *Avoiding $q_{\rm ssb}^{\rm RIKEN}$ havoc?*

Could such a big fundamental quenching (12) generated by quantum anomaly be accommodated in nuclear matter? To address this question let us look at $q_{\rm ssb}$ Eq. (13)

$$q_{\rm ssb} = c_A + (1 - c_A)\Phi^{\beta'}. \tag{13}$$

This equation, governed by a Ward identity,[24] involves two parameters, the constant $c_A$ and the anomalous dimension $\beta'$. Both could in principle depend on density. Now if there were two superallowed GT transitions at two sufficiently different densities involving doubly magic shell nuclei where the ESPM can be applied, the two parameters could be determined and the answer could be given. However up to date such measurements are not available. Even lattice measurement of the $\beta'$ is missing for QCD with 2 flavors. The only information available in the literature I know of is the model calculation of high density phase transitions in the skyrmion model.[47] It was found that for the chiral phase transition to take place at high density, the

*M. Rho*

anomalous dimension $\beta'$ has to be large and positive, say, $\geq 3$. I think this is far-fetched and most likely irrelevant but given the improved RIKEN data, a non-zero $\beta'$ intervening in the $^{100}$Sn decay could not be ruled out.

### 2.4.7. *Prediction with an IR fixed point with QCD-CD*

In the QCD-CD approach,[25] if the two hidden symmetries (HLS and HSS) are considered on the same footing, the $FQ$ can play the role that is played in the GD approach. It is absent in Ref. 25 because the QCD-conformal dilaton has been "integrated out" together with the HLS mesons. In this case, it could be argued as in the case of $\mathcal{N}=1$ SUSY that the anomalous dimension at the IR fixed point vanishes, $\beta'_{IR}=0$. This would render $\Phi^{\beta'}=1$ in (13) for any density and $q_{\rm ssb} \to 1$ independently of density, disagreeing with (12). As emphasized by Zwicky, the "large" dilaton mass which must be related to the explicit quark mass in the QCD Lagrangian could render $\beta$ function run in a complicated way and give important corrections to the anomalous dimension in (13).

It is worth summarizing the key results of this paper in light of how the GSI and RHIC experiments impact on weak processes in nuclear matter.

Although no definite verdict can be reached at present, there are two distinctive possibilities giving two alternative, strikingly different, scenarios. What makes one arrive at these scenarios is the precise matching established between the Landau Fermi liquid fixed-point and the ESPM. If the GSI result is correct, then we will have $q_{\rm ssb}=1$, i.e. no $FQ$, and $g_A^{\rm eff}=1$ together with $\beta'_{IR}=0$. There will be no paradigm change in nuclear physics.

On the other hand, if the RIKEN data are correct, then $q_{\rm ssb} \neq 1$, with significant $FQ$, $g_A^{\rm eff} \approx 1/2$ or $\beta'_{IR} \gg 0$ and the GT amplitude in the neutrinoless $\beta\beta$ process will be strongly suppressed, say, by $<1/16$. The RIKEN result (12) raises a serious challenge both in theory and in experiment. How this effect has remained undetected in nuclear processes studied for so long will pose a mystery. It should have created a havoc in nuclear physics. Whatever the case is, it will bring extremely useful information as to how scale/conformal symmetry manifests in conjunction with the conformal sound velocity in massive compact stars (more on this point below).

## 3. Further Remarks

### 3.1. *Pseudo-conformal symmetry in nuclear matter*

The result obtained in this paper exposes the "old" $g_A$ problem with a totally "new" face. In nuclei, $g_A^{\rm eff} \approx 1$ is captured entirely by nuclear correlations. In highly dense matter as in compact stars, on the other hand, with the same Lagrangian treated in the limit that the re-parametrized field $\langle \chi e^{\frac{i\pi}{f_\pi}} \rangle$ goes to zero, called "dilaton-limit fixed point (DLFP)", which corresponds to the dilaton going massless, $g_A^{\rm eff}$ is also found to approach 1.[16,17] This can be taken as signaling the approach to scale invariance. Since the scale symmetry and chiral symmetry are locked, this limit is equivalent to chiral symmetry restoration manifesting in the Adler–Weisberger-type sum rule. An interesting scenario is that in approaching this limit in conjunction with the vector manifestation, the pion, the dilaton, the $\rho$ (and $a_1$) all go massless, not necessarily at the same density, giving rise to what we might identify with Weinberg's "mended symmetry" multiplets.[48] At an asymptotic high density, perturbative QCD predicts that the sound velocity in neutron stars should become "conformal", $(v_s/c)^2 = 1/3$. Now what is predicted in G$n$EFT is surprising: The conformal sound velocity precociously sets in at a density $n \gtrsim 3n_0$,[17] far below what is usually expected, $\gtrsim 50n_0$. This phenomenon was referred to as "pseudo-conformal". This suggests that $g_A^{\rm eff} \to 1$ at $n \gtrsim 3n_0$. We interpret this as signaling the emergence of the hidden scale symmetry in baryonic medium. This observation is consistent with that in pionless EFT, with all mesons — including the pion — *integrated out*, the *unitarity limit* is applicable in nuclei and compact stars at low density[49] and in $\mathcal{L}_{\psi\sigma HLS}$ with all mesons *integrated in*, the pseudo-conformality is applicable to the physics of compact stars.

While $g_A$ is intrinsically scale-invariant, its "quenching" in nuclear medium depends on the scaling of the dilaton condensate $\Phi$. This is because the nucleon mass scales in medium and the quenching is solely due to strong particle–hole correlations affected by the quasiparticle interactions, not involving chiral-filter unprotected terms. This has a strong implication in the calculation of the Gamow–Teller matrix elements in neutrinoless double beta decays where the momentum transfer can be of $\sim$100 MeV.

### 3.2. $V_{\rm lowk}$-RG to map Landau–Migdal Fermi liquid to extreme single-particle shell model

Let me make a further remark regarding the connection between the Landau–Migdal Fermi-liquid approach and "first principles" many-body approaches in nuclear physics. Such a connection has been anticipated since a long time starting with Walecka's relativistic mean field approach to nuclear many-body problem and the Fermi liquid structure followed by the realization that the LFL fixed point theory can be bridged to the hidden local symmetric and hidden scale symmetric $\chi$EFT. What I consider totally new is that the LFL fixed-point theory, a bona-fide field theory, can be linked to the shell model, ESPM. Now the pertinent question in nuclear physics could be how to accurately calculate the $g_A$ quenching in finite nuclei where the FLFP condition and the ESPM condition cannot be fully exploited. The answer is that although it would be a lot more complicated and indirect one should be able to apply the $V_{\rm lowk}$-RG method as was done in treating the onset of the PC sound speed $(v_s/c)^2 \to 1/3$ in compact-star physics.[26] There it was found that the pseudo-conformality crucially depended on the behavior of the HLS coupling constant approaching the vector-manifestation (VM) fixed point where gauge symmetry becomes unhidden, that is, emergent.

### 3.3. *Remaining mystery*

But this leaves one mystery glaringly unexplained. As shown in $V_{\text{lowk}}$-RG, first in Ref. 26 and reviewed with more rigor recently in Ref. 50, the pseudo-conformal sound speed sets in precociously at a low density with a bump at $\lesssim 3n_0$ and stays conformal through the interior of massive compact stars, if the HLS gauge structure, i.e. vector manifestation, is to emerge at high density, $\gtrsim 25n_0$. This behavior of the EoS results from an intricate interplay of the dilaton and the $\omega$ meson with parity-doubling symmetry, involving a sort of hadron-quark continuity.[51,p] On the other hand, if the HLS gauge structure is to set in at a lower density, say, $\sim 6n_0$, considered by some nuclear theorists to be feasible, then the sound velocity is found to continuously increase, with no peak at $\sim 3n_0$, and go up beyond the conformal velocity. It has been argued[52] that such a sound velocity structure is totally consistent with global star observables described entirely in hadronic variables with no phase transitions or even without hadron-quark continuity.

In a stark contrast, as described in this paper, $g_A$ — modulo $q_{\text{ssb}}$ which is from quantum anomaly — stays *effectively* equal to 1 from light nuclei to heavy nuclei to dilaton fixed point to the restoration of the scale symmetry. What makes the underlying pseudo-conformality manifest so differently in hadronic matter?

## Acknowledgments

I would like to thank R. J. Crewther and L. C. Tunstall for introducing me to their "Genuine Dilaton" model, R. Zwicky for his approach to the IR fixed-point with a vanishing anomalous dimension and K. Yamawaki for his tuition on recent developments on hidden local symmetry.

## ORCID

Mannque Rho  https://orcid.org/0000-0003-1391-1025

---

<sup>p</sup>Although not at odds with any observables available up to date — and very simple, somewhat surprisingly, this scenario remains almost totally ignored in the field.